\documentclass[aps,prl,twocolumn,superscriptaddress,showpacs]{revtex4-1}
\usepackage{amssymb,amsmath,amsfonts,graphicx,epsf}

\begin{document}


\title{Magnetic Discontinuities in Magnetohydrodynamic Turbulence and in the Solar Wind}


\author{Vladimir Zhdankin}
\author{Stanislav Boldyrev}
\affiliation{Department of Physics, University of Wisconsin, 1150 University Avenue, Madison, Wisconsin 53706, USA}
\author{Joanne Mason}
\affiliation{Department of Astronomy and Astrophysics, University of Chicago, 5640 South Ellis Avenue, Chicago, Illinois 60637, USA}
\author{Jean Carlos Perez}
\affiliation{Department of Physics, University of Wisconsin, 1150 University Avenue, Madison, Wisconsin 53706, USA}
\affiliation{Space Science Center, University of New Hampshire, Durham, New Hampshire 03824, USA}


\date{\today}

\begin{abstract}
Recent measurements of solar wind turbulence report the presence of intermittent, exponentially distributed angular discontinuities in the magnetic field. In this Letter, we study whether such discontinuities can be produced by magnetohydrodynamic (MHD) turbulence. We detect the discontinuities by measuring the fluctuations of the magnetic field direction, $\Delta\theta$, across fixed spatial increments $\Delta x$ in direct numerical simulations of MHD turbulence with an imposed uniform guide field $B_0$. A large region of the probability density function (pdf) for $\Delta\theta$ is found to follow an exponential decay, proportional to $\exp(-\Delta\theta/\theta_*)$, with characteristic angle $\theta_* \approx (14^\circ) (b_{\text{rms}}/B_0 )^{0.65}$ for a broad range of guide-field strengths. We find that discontinuities observed in the solar wind can be reproduced by MHD turbulence with reasonable ratios of $b_{\text{rms}}/B_0$. We also observe an excess of small angular discontinuities when $\Delta x$ becomes small, possibly indicating an increasing statistical significance of dissipation-scale structures. The structure of the pdf in this case closely resembles the two-population pdf seen in the solar wind. We thus propose that strong discontinuities are associated with inertial-range MHD turbulence, while weak discontinuities emerge from dissipation-range turbulence. In addition, we find that the structure functions of the magnetic field direction exhibit anomalous scaling exponents, which indicates the existence of intermittent structures.   
\end{abstract}

\pacs{95.30.Qd, 96.50.Ci, 96.50.Tf}

\maketitle


{\em Introduction.}---Over the past several decades, multiple spacecraft have measured the fluctuations of the magnetic and velocity fields in the solar wind \cite{horbury2005,ness2001}. This has provided a wealth of data from which to test plasma models such as magnetohydrodynamic (MHD) turbulence \cite{goldstein1999,bruno2005}. Among other things, this allows one to study the intermittency of MHD and dissipative structures such as current sheets. However, the question remains of whether the solar wind can be adequately described by MHD turbulence, and whether different types of structures can be distinguished from the data. 

The existence of intermittent structures in a plasma is implied by abrupt changes in magnetic field directions. Bruno et al. \cite{bruno2001} studied this feature in solar wind data from the Helios 2 spacecraft by performing a minimum variance analysis on the magnetic field vector. It was found that there are times when the magnetic field undergoes large changes in direction, implying that the solar wind contains intermittent structures. It was proposed that these strong fluctuations mark the boundaries of flux tubes that originate in the Sun and are passively advected by the solar wind.

Later, Borovsky \cite{borovsky2008} used data from the ACE spacecraft to examine the probability density function (pdf) of angular shift in the magnetic field, given by $\Delta\theta = \cos^{-1}{( \boldsymbol{B}_1 \cdot \boldsymbol{B}_2 / | \boldsymbol{B}_1 | | \boldsymbol{B}_2 | )}$ where $\boldsymbol{B}_1$ and $\boldsymbol{B}_2$ are measurements of the magnetic field taken at two different times. Two populations of magnetic discontinuities were discerned. The first population consists of strong discontinuities at $30^\circ < \Delta \theta < 170^\circ$ with an exponentially decaying pdf proportional to $\exp(-\Delta\theta/24.4^\circ )$. The second population consists of weak fluctuations at $5^\circ < \Delta \theta < 30^\circ$ which can be fit by exp($-\Delta\theta/9.4^\circ$). Miao et al. \cite{miao2011} performed a similar analysis on slow wind data from the Ulysses spacecraft. In their case, the first population spanned approximately $50^\circ < \Delta \theta <160^\circ$ and was proportional to exp($-\Delta\theta/30.0^\circ$), while the second population spanned $30^\circ < \Delta \theta < 50^\circ$ and was proportional to exp($-\Delta\theta/18.6^\circ$), broadly agreeing with Borovsky's result.  In both reports, the strong discontinuities were interpreted to come from coronal flux tube walls, while the weak discontinuities were assumed to be turbulent fluctuations. 

An alternative hypothesis is that the magnetic discontinuities in the solar wind are predominantly generated by nonlinear interactions \cite{li2008, neugebauer2010, vasquez2007}. In this case, the discontinuities evolve dynamically as the solar wind expands away from the Sun. It is known that current sheets form spontaneously in MHD turbulence, providing a natural source of discontinuities.

These studies raise a principle question of why the pdfs of angular shifts have exponential laws (regardless of their origin), and what determines the typical angular discontinuities characterizing the two observed scalings. Although the statistical properties of magnetic fields in MHD turbulence have been studied before \cite{zhou2004,uritsky2010,greco2008}, the statistical properties of magnetic discontinuities have been addressed to a lesser extent \cite{greco2009,greco2010}. In particular, the statistical properties of angular shifts have not been studied in numerical simulations of MHD turbulence. In this Letter, we investigate the statistical properties of magnetic discontinuities in direct numerical simulations of MHD turbulence with an imposed uniform magnetic field $B_0$. We use simulations with several choices of $B_0$ that span the transition from turbulence with a weak mean field to turbulence with a strong mean field. We find that the pdfs of $\Delta\theta$ all contain a region well fit by $P(\Delta \theta)\propto \exp(-\Delta\theta/\theta_*)$, with a scaling that depends on the background field strength as $\theta_* \approx (14^\circ)\,(b_{\text{rms}} / B_0 )^{0.65}$. The pdf of strong discontinuities in Borovsky's and Miao et al.'s studies could then be explained by MHD turbulence with reasonable fluctuations to guide-field ratios for the solar wind: the first population in Borovsky's results is consistent with our results for $b_{\text{rms}}/B_0 \sim 2.4$, and Miao et al.'s result is consistent with $b_{\text{rms}}/B_0 \sim 3.2$.  

We also find that when the spatial increment $\Delta x$ becomes small enough, the pdf exhibits an excess of small angular discontinuities, which closely resembles the ``second'' population in the solar wind observations. We conjecture that this population indicates the increasing contribution of structures near the dissipative scale, where the nature of turbulence changes.

 One can then envision the following explanation of the solar wind observations in which both populations of discontinuities arise from MHD turbulence. According to our analysis, the strong discontinuities may be produced by discontinuities in MHD turbulence with $b_{\text{rms}} > B_0$, which is typical of the solar wind. These discontinuities will be observed regardless of whether any flux tubes are advected by the solar wind. The population of weak discontinuities, on the other hand, may be produced by a different kind of turbulence, which still generates an approximately exponential pdf but with a different characteristic angle. In particular, this population may be associated with near-dissipation-scale turbulence. If so, this population emerges only when the interval between measurements is small enough. 

This picture is attractive because it explains the origin and the exponential pdf of large angular discontinuities as well as the small ones. This model for the solar wind magnetic discontinuities can be complementary or possibly alternative to Borovsky's flux tube hypothesis.

{\em Method.}---The incompressible MHD equations can be written as
\begin{eqnarray}
\partial_t \boldsymbol{v} + (\boldsymbol{v} \cdot \nabla) \boldsymbol{v} &=& - \nabla p + (\nabla \times \boldsymbol{B}) \times \boldsymbol{B} + \nu \nabla^2 \boldsymbol{v} + \boldsymbol{f}, \nonumber \\
\partial_t \boldsymbol{B} &=& \nabla \times (\boldsymbol{v} \times \boldsymbol{B}) + \eta \nabla^2 \boldsymbol{B}, \nonumber  \\
\nabla \cdot \boldsymbol{v} &=& 0, \nonumber \\
\nabla \cdot \boldsymbol{B} &=& 0,
\end{eqnarray}
where $\boldsymbol{v}(\boldsymbol{x},t)$ is the plasma velocity, $\boldsymbol{B}(\boldsymbol{x},t)$ is the magnetic field, $p$ is the pressure, and $\boldsymbol{f}(\boldsymbol{x},t)$ is the external forcing. We take the viscosity $\nu$ and resistivity $\eta$ to be equal.

The angular shift in magnetic field between two spacecraft measurements is the angle between $\boldsymbol{B}(t)$ and $\boldsymbol{B}(t+\Delta t)$, where $\boldsymbol{B}(t)$ is the magnetic field vector at time $t$ and $\Delta t$ is the time increment. Instead of time increments, we use a spatial increment in our analysis, which can be approximately related to the time increment by $\Delta x \approx V_{SW} \Delta t$, where $V_{SW}$ is the solar wind velocity. The angular shift in magnetic field between two points $P_1 = (x,y,z)$ and $P_2 = (x+\Delta x,y,z)$ is then given by
\begin{align}
\Delta\theta = \cos^{-1}{ \left( \frac{\boldsymbol{B}(x,y,z) \cdot \boldsymbol{B}(x+\Delta x,y,z)}{|\boldsymbol{B}(x,y,z)| |\boldsymbol{B}(x+\Delta x,y,z)| } \right) }.
\end{align}

We analyze data from simulations of driven incompressible MHD with five different guide-field strengths: $B_0 \in \{0.25, 0.5, 1, 5, 10\}$, in comparison to the root mean square average perpendicular fluctuations of $b_\text{\text{rms}} \sim 1.3$. The simulations solve the full MHD equations with a Reynolds number $\text{Re} \approx 2200$ and conditions similar to Ref. \cite{mason2006}. The simulations have a resolution of $512^3$.

\begin{figure}[!t]
 \includegraphics[width=\columnwidth]{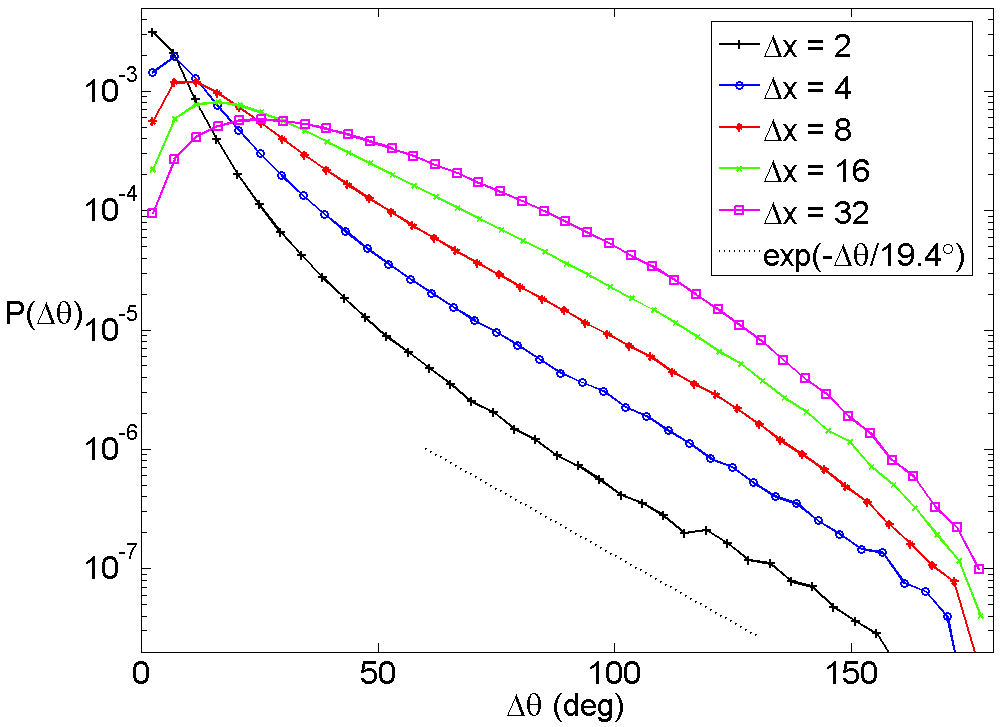}
\caption{The pdfs for angular shifts in magnetic field for MHD turbulence with fixed guide field $B_0 = 1$ and measurement increments in the range $2 \le \Delta x \le 32$. Each pdf has a region that is approximately an exponential decay, with a characteristic angle that is independent of $\Delta x$. Note that for $\Delta x$ near the dissipation scale, small angular shifts become more abundant than expected from the exponential tail. \label{fig1}}
 \end{figure}

\begin{figure}[!t]
 \includegraphics[width=\columnwidth]{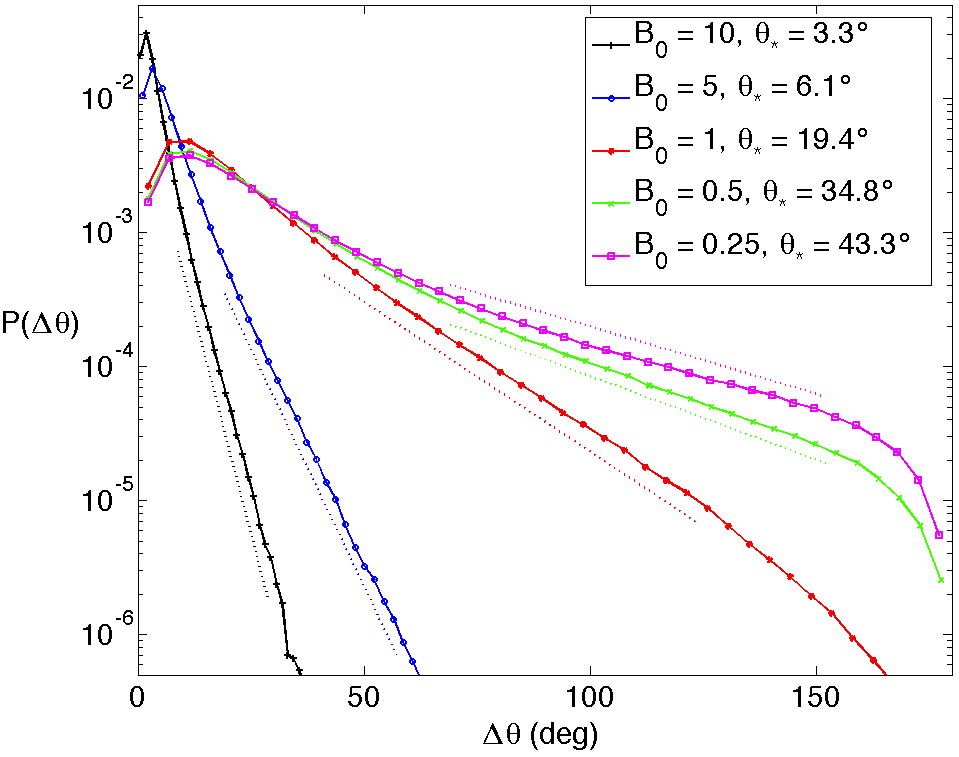}
 \caption{The pdfs for angular shifts in magnetic field with fixed $\Delta x = 8$ and several background magnetic fields (where $b_\text{\text{rms}} \sim 1.3$). Each pdf has a region where it takes the form of exp($-\Delta\theta/\theta_*$), where $\theta_*$ depends on $B_0$. \label{fig2}}.
 \end{figure}

For each case of $B_0$, multiple snapshots of the simulation during steady-state are analyzed. In each snapshot, we measure $\Delta\theta$ on a line of points separated by $\Delta x$ in the $x$-direction (this is perpendicular to the guide field, but the results in the parallel direction are similar). Spatial increments are taken in the range $2 \le \Delta x \le 32$, where $\Delta x = 1$ is equal to the mesh spacing. This is repeated along other lines in the snapshot with a spacing of $\Delta y$ and $\Delta z$ grid points, where $\Delta y$ and $\Delta z$ are varied to give enough statistics. For each choice of $\Delta x$ and $B_0$, we measured $\Delta\theta$ at approximately $3.5 \times 10^6$ data points.\\

{\em Results.}---The pdf for angular shifts in magnetic field is shown in Fig. \ref{fig1} for fixed guide field of $B_0 = 1$ and spatial separations in the range $2 \le \Delta x \le 32$. The axes of the plot are log-linear, so a straight line represents an exponential function. Each pdf has a region that is fit by exp($-\Delta\theta/\theta_*$), where the characteristic angle $\theta_* = 19.4^\circ$ is independent of $\Delta x$. For small spatial separations ($\Delta x \le 4$), the exponential region is limited to the tail of the pdf, and small angular shifts are more abundant than expected from the exponential fit. This deviation from an exponential can be attributed to dissipative effects, and provides a natural explanation for a distinct population of weak angular discontinuities. 

Next, we consider how the pdf for angular shifts changes when the guide field is varied. This is shown in Fig. \ref{fig2} for fixed $\Delta x = 8$ and varying $B_0$. Each case has an exponential region, but with characteristic angle $\theta_*$ decreasing with stronger guide fields. The value of $\theta_*$ is acquired by measuring the slope in the exponential region with a least-squares fit. The result is that $B_0 = [0.25, 0.5, 1, 5, 10]$ correspond to $\theta_* = [43.3^\circ, 34.8^\circ, 19.4^\circ, 6.1^\circ, 3.3^\circ]$, respectively.

\begin{figure}[!t]
 \includegraphics[width=\columnwidth]{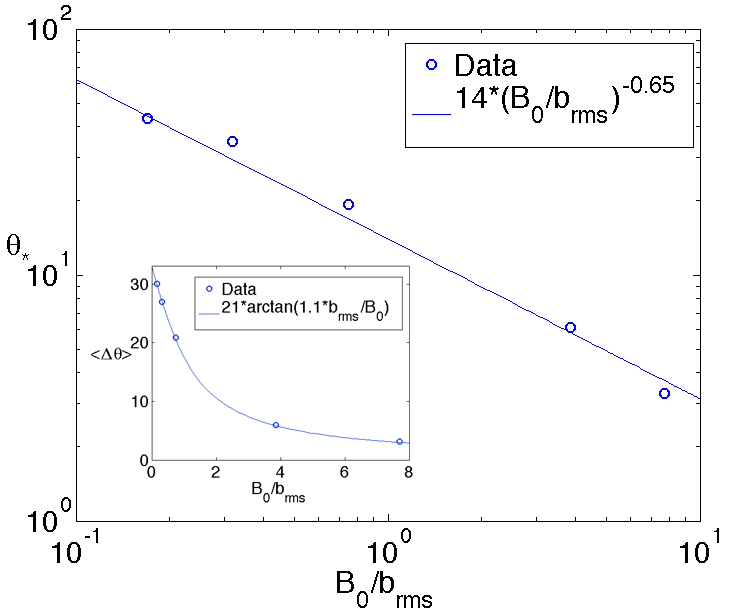}
 \caption{The characteristic angle measured in the exponential part of the pdf of $\Delta\theta$, denoted $\theta_*$, is found to have a power law dependence on $B_0/b_{\text{rms}}$. The fit shown is $\theta_* \approx (14^\circ) (b_{\text{rms}}/B_0)^{0.65}$. It is interesting to note that the mean angular shift, $\langle \Delta\theta \rangle$, instead has an arctan dependence, shown in the inset. \label{fig3}} 
 \end{figure}

A functional form of the relation between $\theta_*$ and $B_0/b_{\text{rms}}$ is found by making a fit to our numerical measurements. The best fit is a power law (Fig. \ref{fig3}), given by 
\begin{align}
\theta_* \approx (14^\circ)\, (b_{\text{rms}}/B_0)^{0.65}. \label{angle_eq}
\end{align}

It is interesting to point out that the mean angular shift obeys a different empirical relation (obtained by fitting the simulation data),
\begin{align}
\langle \Delta\theta \rangle = \int \Delta\theta P(\Delta\theta) d\Delta\theta \propto  \arctan{\left(\frac{1.1 b_{\text{rms}}}{B_0}\right)} ,
\end{align}
which is shown in the inset of Fig. \ref{fig3}. This reflects the fact that a significant contribution to the average comes not from the exponentially declining tail of the probability density function, but rather from its bulk part that has a different scaling. Hence, the mean value of $\Delta\theta$ is determined by fluctuations outside of the exponential region. In support of this view, we note that the location of the maximum in the function $\Delta\theta P(\Delta\theta)$ indeed has an arctan dependence on $B_0$. This result reflects the fact that a strong background magnetic field prevents large deviations of the magnetic field vector from $\boldsymbol{B} = B_0 \boldsymbol{\hat{z}}$.

In addition to pdfs, another statistical tool used to study intermittency is the spatial scaling of structure functions \cite{marsch1997, padoan2004}. While the anomalous scaling of various structure functions has been observed in fluid and in hydromagnetic turbulence, the intermittency of magnetic angular discontinuities has not been studied before. We define the structure function of order $n$ for magnetic field direction to be
\begin{align}
S_\theta^n (\Delta x) &= \langle | \Delta\theta (\Delta x) | ^n \rangle ,
\end{align}
where the brackets $\langle \cdots \rangle$ indicate the spatial average of the enclosed quantity. We assume that structure functions have the form
\begin{align}
S_\theta^n (\Delta x) &\propto (\Delta x)^{\zeta_n},
\end{align}
where $\zeta_n$ are the exponents of the structure functions.

\begin{figure}[t!]
 \includegraphics[width=\columnwidth]{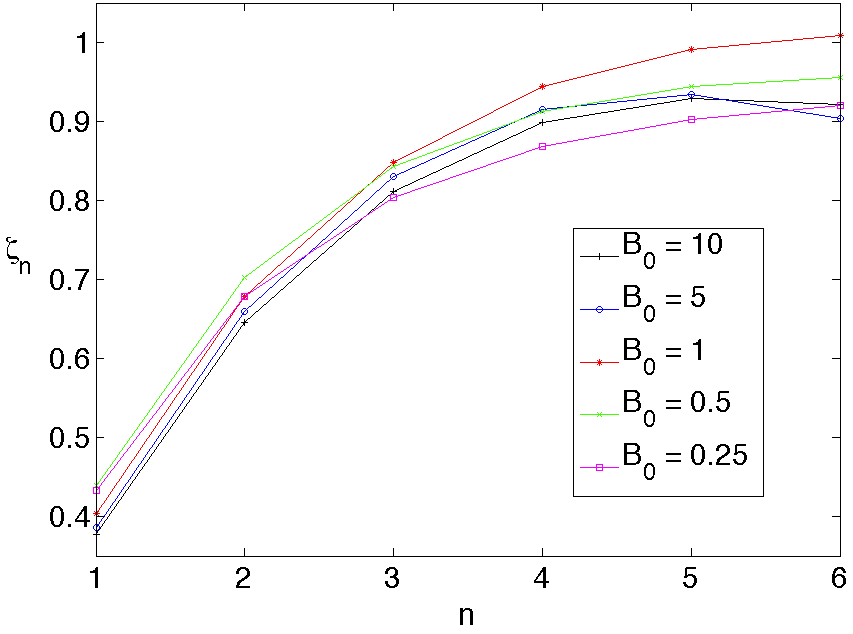}
 \caption{The structure function exponents, $\zeta_n$, associated with the structure functions of magnetic field direction, $S_\theta^n$. The deviation from a straight line implies the existence of intermittent structures \label{fig4}.}
 \end{figure}

We determine $\zeta_n$ by measuring the slope of $S_\theta^n$ versus $\Delta x$ on a log-log plot using a least-squares fit. As shown in Fig. \ref{fig4}, the relationship between $\zeta_n$ and $n$ is not linear, which indicates anomalous scaling due to intermittency. Flattening around $\zeta_n\approx 1$ implies the presence of shocklike magnetic angular discontinuities, that is, current sheets \cite{frisch1995, muller2003}. It is important to note that measurements of structure functions above order 5 or 6 in a finite series of data may be contaminated by contributions from rare events \cite{DudokdeWit2004}, hence the rightmost values in Fig. \ref{fig4} could be unreliable.

In addition to angular shifts in the magnetic field, one may consider studying discontinuities in velocity field. However, the angular shift in velocity is not invariant under a Galilean transformation, so it has an ambiguous physical interpretation. Instead, consider the velocity jump,
\begin{align}
\Delta v = | \boldsymbol{v}(x + \Delta x,y,z) - \boldsymbol{v}(x,y,z) |
\end{align}
which also has a pdf with an exponential region (not shown) \cite{marsch1997}. A scatterplot not shown here reveals that $\Delta\theta$ is correlated with $\Delta v$ at large values, as long as $\Delta x$ is chosen to be small ($\Delta x \le 8$).

Interestingly, a similar correlation was found in solar wind data for fractional velocity jump $\Delta v / v$ instead of velocity jump, where $v$ is approximately constant (the solar wind velocity) \cite{borovsky2008}. The correlation between $\Delta v / v$ and $\Delta\theta$ at large values of $\Delta\theta$ was attributed to the crossing of flux tube walls, but our results suggest that this can be explained by current sheets produced from turbulence.  \\

{\em Conclusions.}---We have addressed to what extent incompressible MHD turbulence can describe the magnetic discontinuities in the solar wind. This was done by studying the statistical properties of angular shifts in turbulent MHD simulations with varying strengths of guide field. We found that the pdf of angular shifts has a region of exponential decay, $P(\Delta \theta)\propto \exp(-\Delta \theta/\theta_*)$, in agreement with the solar wind. We found that the associated characteristic angle is independent of spatial increment $\Delta x$, but depends on the guide-field strength as $\theta_* \approx (14^\circ) (b_{\text{rms}}/B_0 )^{0.65}$.

Our results have implications for the discontinuities observed in the solar wind. The exponential pdf of strong discontinuities in the solar wind is consistent with MHD turbulence for values of $b_{\text{rms}}/B_0$ that are typical of the solar wind. Additionally, our simulations show an excess of weak discontinuities when spatial increments are small, which can be interpreted as a second population. Turbulence can therefore qualitatively reproduce the solar wind observations, with the strong discontinuities associated with inertial-scale turbulence, and the weak discontinuities with near-dissipation-scale turbulence. MHD turbulence can then be considered a complementary or alternative explanation of the discontinuities in the solar wind. Further support for this picture comes from the correlation between large velocity jumps and angular shifts in magnetic field observed in our simulations.

We also found that the structure functions of the magnetic field direction provide a means to study intermittency. The observations of magnetic discontinuities in the solar wind, and their connection to MHD turbulence, motivate further study. A more detailed analysis of these results will be presented in a future paper.

This work was supported   
by the US DoE Awards DE-FG02-07ER54932, DE-SC0003888, DE-SC0001794, the NSF Grant PHY-0903872, the NSF/DOE Grant AGS-1003451, and the NSF Center for Magnetic Self-organization in Laboratory and Astrophysical Plasmas at U. Wisconsin-Madison and the University of Chicago. High Performance Computing resources were
provided by the Texas Advanced Computing Center (TACC) at the
University of Texas at Austin under the NSF-Teragrid Project
TG-PHY080013N and by the National Institute for Computational Sciences. 

\bibliography{refs_angles}

\end{document}